\documentclass[a4paper,11pt]{article}
\usepackage{pos}
\usepackage{xspace}
\usepackage[capitalize]{cleveref}
\usepackage{enumitem,wrapfig}
\usepackage{xcolor}\definecolor{ForestGreen}{rgb}{0.13,0.55,0.13}\definecolor{Cerulean}{rgb}{0,0.48,0.65}
\usepackage[percent]{overpic}
\title{Reconstructing Air-Shower Observables using a Universality-Based Model at the Pierre Auger Observatory}
\ShortTitle{Shower Universality}

\author[abc]{Maximilian Stadelmaier}  % <-- please don't change authors since this was discussed in the icrc comm and this has been decided

\affiliation[a]{Istituto Nazionale di Fisica Nucleare \href{https://www.mi.infn.it/}{\raisebox{-0.2\height}{\includegraphics[height=2.5ex]{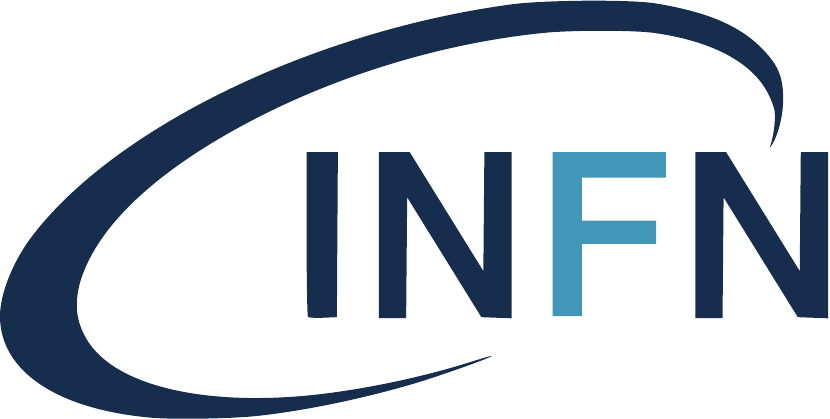}}}, Sezione Milano, Milano, Italy}
\affiliation[b]{Dipartimento di Fisica, Universit\`a degli Studi di Milano \href{https://fisica.unimi.it/}{\raisebox{-0.25\height}{\includegraphics[height=1.2em]{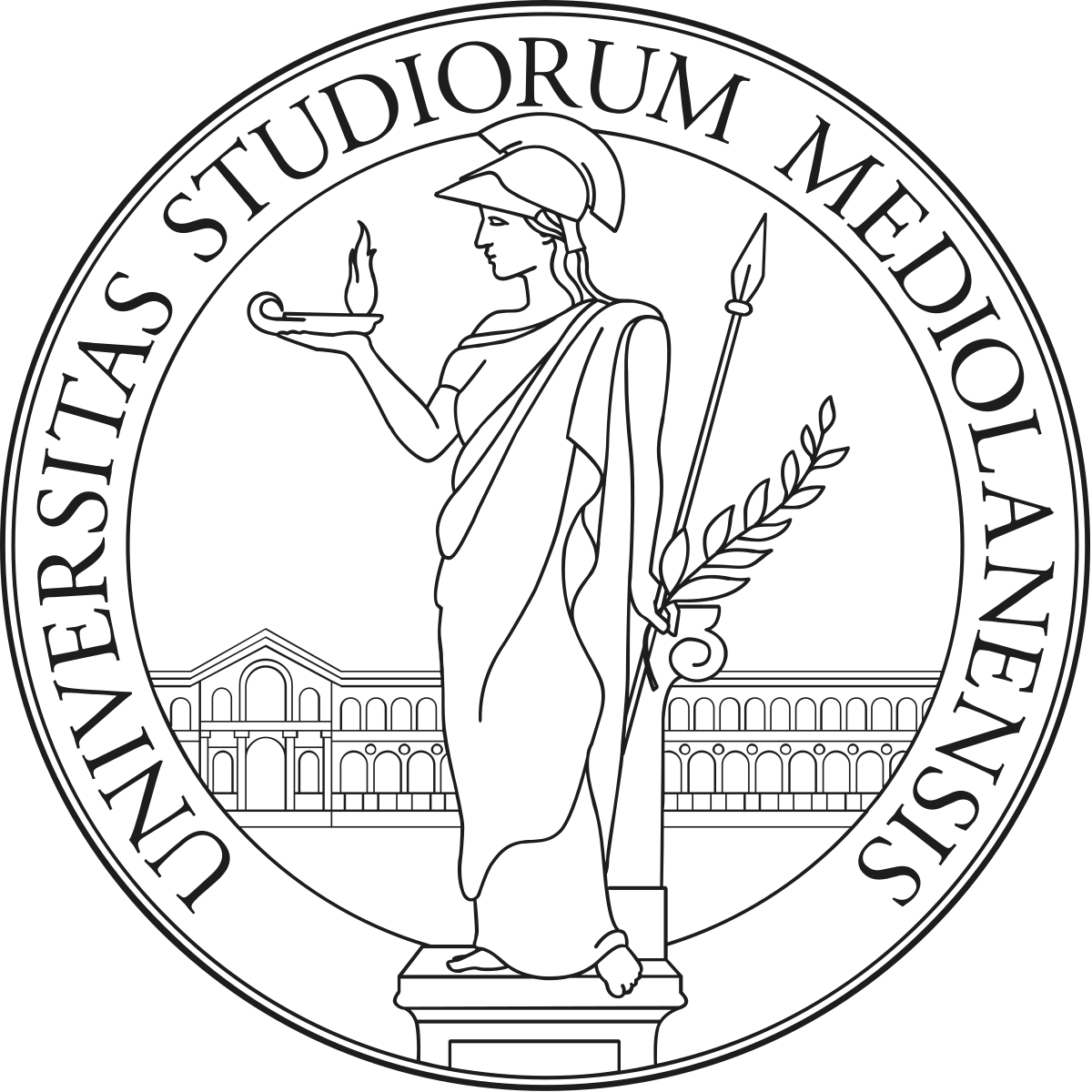}}}, Milano, Italy}
\affiliation[c]{Institute for Astroparticle Physics, Karlsruhe Institute of Technology \href{https://www.iap.kit.edu/}{\includegraphics[height=1.55ex]{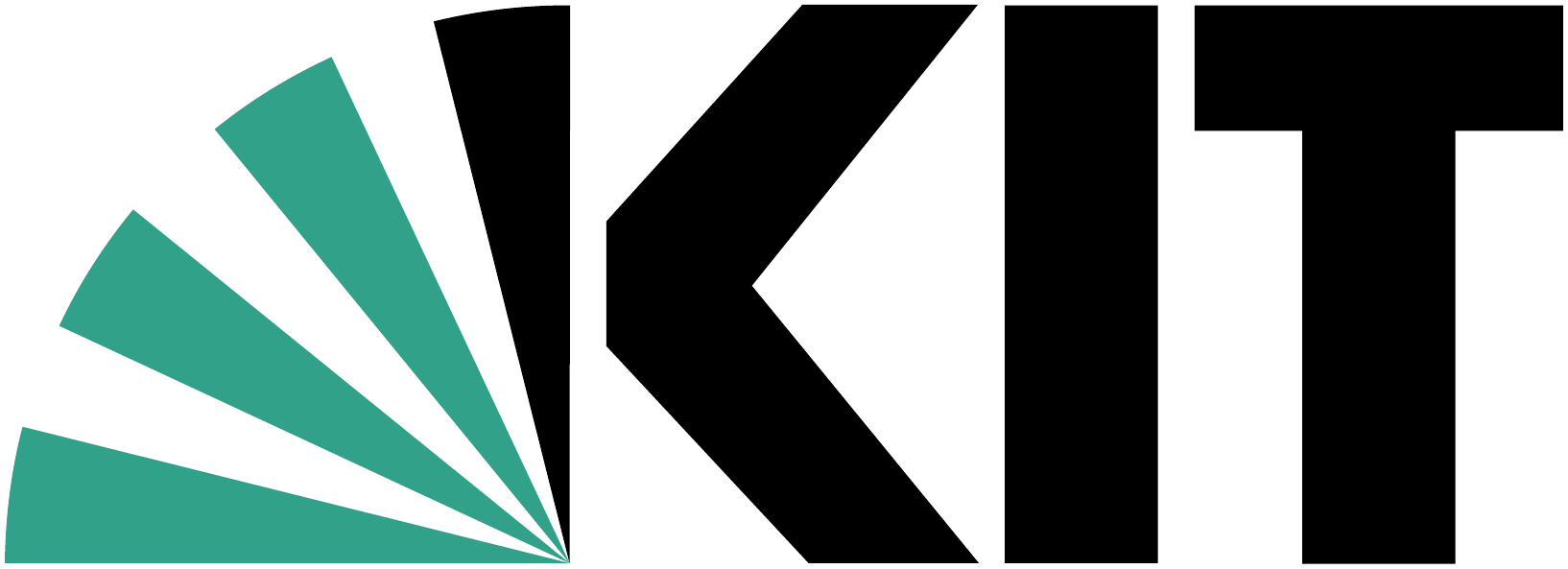}}, Karlsruhe, Germany}

\onbehalf{for the Pierre Auger Collaboration$^d$}
\affiliation[d]{Observatorio Pierre Auger, Av.\ San Mart{\'\i}n Norte 304, 5613 Malarg\"ue, Argentina\\
Full author list: {\rm\url{https://www.auger.org/archive/authors_icrc_2025.html}}}

\emailAdd{spokespersons@auger.org}

\abstract{
Based on solutions of the cascade equations, the \emph{air-shower universality} is a framework that for all air showers with the same energy, zenith angle, depth of shower maximum, and muon number predicts the same longitudinal, lateral, and energy distributions of electromagnetic shower particles.
We employ a universality-based model of shower development that incorporates hadronic particle components to reconstruct observables from extensive air showers produced by ultra-high-energy cosmic rays.
The model can estimate key parameters, such as the depth of the shower maximum and the number of muons at the event level.
We discuss the performance of the reconstruction algorithm using both air-shower simulations, and preliminary results obtained from the Phase-I data of the Pierre Auger Observatory.
}

\ConferenceLogo{PoS_ICRC2025_logo}

\FullConference{%
39th International Cosmic Ray Conference (ICRC2025)\\
15 -- 24 July 2025\\
Geneva, Switzerland}

\def\Offline{\mbox{$\overline{\textrm{Off}}$\hspace{.05em}\protect\raisebox{.4ex}{$\protect\underline{\textrm{line}}$}}\xspace}

\def\Xmax{X_\text{max}}

\def\avg#1{\langle{#1}\rangle}
\def\gcm{\text{g/cm$^2$}}

\def\orcid#1{\href{https://orcid.org/#1}{\includegraphics[height=1.55ex]{orcid-ID}}}

\def\un2{\texttt{Universality\,II}}
\def\m{\text{m}}
\def\eV{\text{eV}}

\def\EeV{\text{EeV}}

\def\red#1{{\color{red}#1}}

\begin{document}
\maketitle

\section{Introduction}
\label{sec:intro}

The origin and nature of ultrahigh-energy cosmic rays (UHECRs) is an open problem of modern Astroparticle Physics.
In the last decades, the Telescope Array~\cite{TelescopeArray:2008toq} and the Pierre Auger Observatory~\cite{PierreAuger:2015eyc} were established to detect cosmic rays at the highest energies, up to 100\,EeV and beyond, and to resolve conflicting results of previous experiments.
According to the recent results of the Pierre Auger Observatory, the flux of cosmic rays at the highest energies is composed of an extra-galactic component~\cite{Dipole:2017}, which is of a mixed-mass composition~\cite{PierreAuger:2014gko,PierreAuger:2016qzj}, and which
%shows yet unexplained spectral features~\cite{PierreAuger:2020kuy} such as a strong suppression of the flux beyond about 50\,EeV.
shows spectral features~\cite{PierreAuger:2020kuy} such as a strong suppression of the flux beyond about 50\,EeV, for which several different possible explanations have been hypothesized.

Ground-based cosmic ray experiments rely on the detection of the air-shower phenomenon.
Air showers occur when very energetic particles enter the Earth's atmosphere, which acts as a calorimeter in which the energy of the primary cosmic ray is converted into cascades of secondary particles that reach up to several kilometers in diameter and comprise approximately $10^9$ particles per $\EeV$ of the primary energy.
They can be detected directly in clear moonless nights by observing the fluorescence light produced by the excitation of nitrogen in the atmosphere, or indirectly by recording the particles that reach the ground.
Air-shower profiles, observed as fluorescence light, yield valuable information about the primary cosmic rays.
The total fluorescence light emitted is a reliable proxy for the number of particles produced in the air shower, and thus for the energy of the primary; and the development of the air-shower profile through the atmosphere, especially the depth $X_\text{max}$ at which it appears the brightest (the \emph{shower maximum}), yields information about the type of the primary particle.
Being restricted to nights with optimal atmospheric and light conditions, however, the direct detection of shower profiles is only feasible ${\sim}15\%$ of the time.
Surface detectors, which are operational up to 100\% of the time, do not directly record the air-shower development through the atmosphere, but measure the secondary shower particles that reach the ground in terms of the total particle density as well as the arrival time of the particles.
It has been demonstrated that the depth of the shower maximum $X_\text{max}$ can be inferred empirically from the temporal information of the particles reaching the ground~\cite{PierreAuger:2017tlx}.
In this way, information about the shower development is accessible at the highest energies, where the flux of UHECRs is too low to collect a sufficient number of events during the operational time of fluorescence detectors.
In this work, we present $X_\text{max}$ of UHECRs recorded by the surface detector of the Pierre Auger Observatory at primary energies above $4\,\EeV$, estimated using a novel method that is based on the established idea of an air-shower universality~\cite{Hillas:1982vn,Lipari:2008td,NERLING2006421}.
The method makes use of a physically motivated model of the particle densities in air showers and can be extended to extract also other air-shower observables.

\section{The Pierre Auger Observatory}
\label{sec:auger}

The Pierre Auger Observatory is the largest cosmic ray detector in the world.
Up to this date it has collected an unprecedented amount of data and exposure during its ${\sim}20$ years of operation.
The observatory is located on the plateau of the Argentinean Pampa Amarilla at an average altitude of 1400\,\m above sea level.
It is equipped with a set of fluorescence detector (FD) telescopes as well as with a 3000\,km$^2$ surface detector (SD) array~\cite{PierreAuger:2009esk,PierreAuger:2007kus}.
Its hybrid detector setup allows for an absolute calibration of the SD using the FD telescopes.
The main SD array is comprised of ${\sim}1600$ water-Cherenkov detectors (WCDs).
The WCDs are each filled with 12\,\text{tons} of purified water and collect the Cherenkov light emitted by through-going air-shower particles using three photo-multipliers.
The signals are digitized and sampled in 25\,ns time bins.
The detectors are calibrated in terms of VEM, which is the most probable signal of a vertical through-going atmospheric muon~\cite{PierreAuger:2005znw}.
For an overview of the detector operation see~\cite{PierreAuger:2015eyc}.

The surface detector is fully efficient to detect cosmic rays above a primary energy of $E_0=10^{18.5}\,\text{eV}=3$\,EeV.
Approximately 50 UHECRs are detected each day above full efficiency and within a zenith angle\footnote{The Pierre Auger SD is capable of also detecting cosmic rays with arrival directions within $60^\circ<\theta<80^\circ$, however, due to larger asymmetries arising from the geomagnetic field a special reconstruction technique is applied for such inclined events~\cite{PierreAuger:2014jss}.} of $\theta\lesssim60^\circ$~\cite{PierreAuger:2020yab}, and approximately 25 UHECRs are recorded each day with energies above 4\,EeV.

\section{Measurement of the Mass Composition of Cosmic Rays}
\label{sec:mass}

The mass composition of UHECRs cannot be measured directly, but only through proxy observables from air-shower measurements.
The atmospheric depth $\Xmax$, at which the shower reaches its maximum, is directly linked to the nuclear mass $A$ of the primary cosmic ray~\cite{MATTHEWS2005387}.
On average, lighter primary particles of a given energy produce deeper showers, while heavier have shallower development.
At the same time, the number of muons in air showers initiated by heavy nuclei is enhanced relatively to the lighter primaries.

Both the number of muons and $\Xmax$ can be accurately measured using the hybrid detector of the Pierre Auger Observatory~\cite{PierreAuger:2014ucz,PierreAuger:2014gko,PierreAuger:2023vcr}.
Using surface-detector data only, however, this is a challenging task.
The surface detector alone has no direct access to information about the air-shower profile, and thus the calorimetric energy deposit as well as the shower development cannot be measured.
The energy $E_0$ of the primary particle and $\Xmax$, however, can be estimated from the shower footprint on the ground and from the temporal distribution of the particles in the detectors, respectively~\cite{PierreAuger:2020yab,PierreAuger:2017tlx,PierreAuger:2024nzw}.
The number of muons, as well, can be estimated from the footprint of the shower at the ground, but, depending on how the total energy estimate is obtained, can be significantly biased with respect to the expectations for different primary particles.
An unbiased estimate of the muon number can easily be attained from the shower footprint if an independent energy estimate such as from the FD is available.

\section{The Universality Shower Model}
\label{sec:univ}

The model of particle densities used in this work is based on \emph{air-shower universality}~\cite{Hillas:1982vn,NERLING2006421,Lipari:2008td,Stadelmaier:2024pae}
according to which the expected distribution of particles at the ground can be accurately described as a function of the primary energy $E_0$ of the UHECR, the depth $X_\text{max}$ of the shower maximum, the relative number of muons $R_\upmu$, and the event geometry.
The model is parametrized using detector-response simulations of the surface detectors of the Pierre Auger Observatory, produced with the \Offline software framework using \textsc{Corsika} showers generated with the \textsc{Epos-LHC} model of hadronic interactions~\cite{Argiro:2007qg,Heck:1998vt,Pierog:2013ria}.

In the model, the particle content of the shower is divided into different subspecies (components), which are treated separately.
In this way, even showers with a large hadronic contribution to the particle content can be described universally.
Both the expected lateral and longitudinal development of the shower are parametrized for all particle components as a function of $E_0$, $X_\text{max}$, and $R_\upmu$ of muons created in the cascade; additionally, the temporal distribution of the particles at the ground is parametrized for each component as a function of $X_\text{max}$ and the event geometry.
See Ref.~\cite{Stadelmaier:2024pae} for a detailed description of the model itself and the reconstruction mechanism.

\section{Golden Hybrid Data: Number of Muons}
\label{sec:rmu}

\begin{figure}
    \centering
    \def\h{0.36}
    \begin{overpic}[height=\h\textwidth]{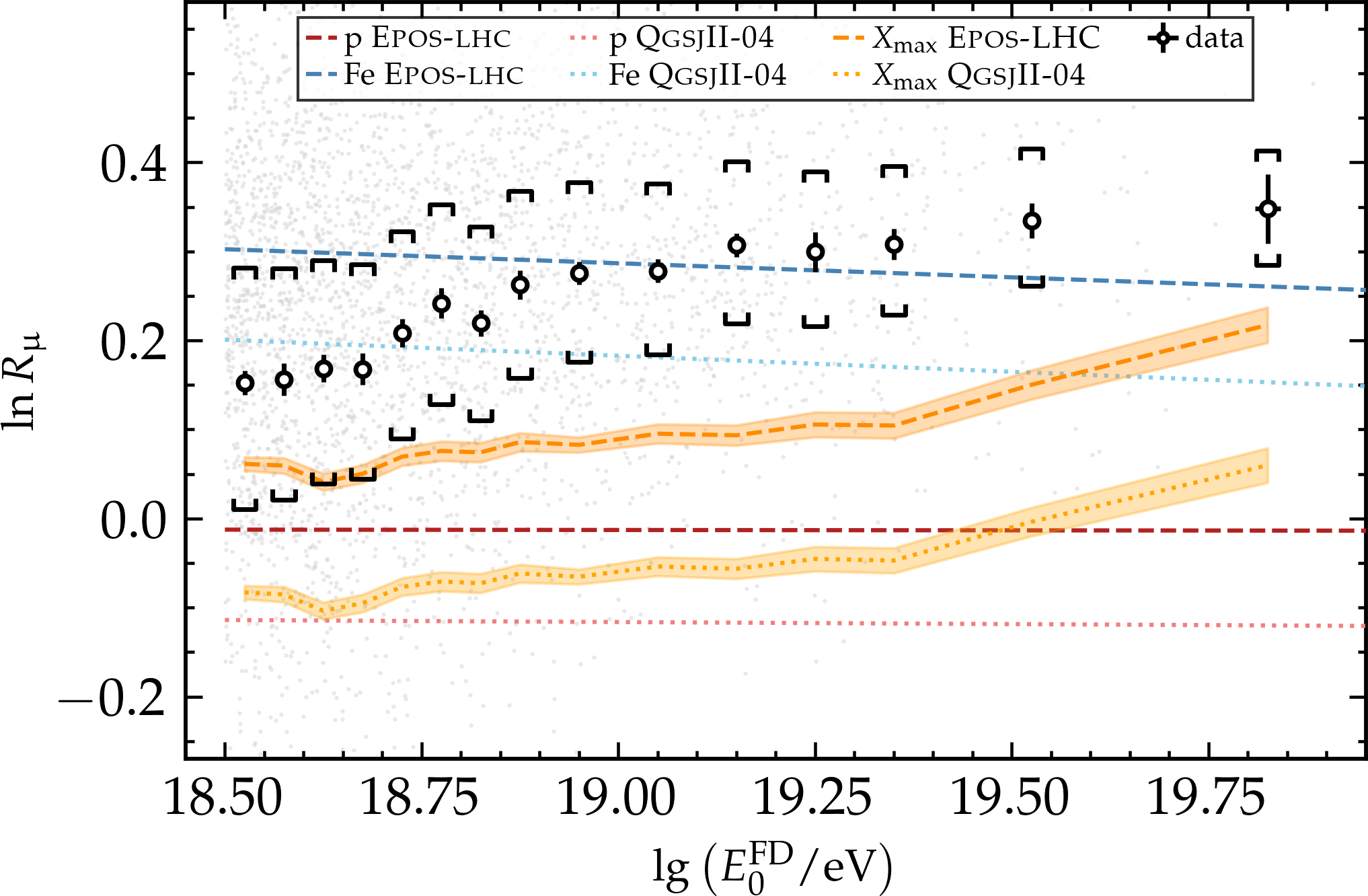}
    \put(74,14){\red{Preliminary}}
    \end{overpic}\hfill
    \begin{overpic}[height=\h\textwidth]{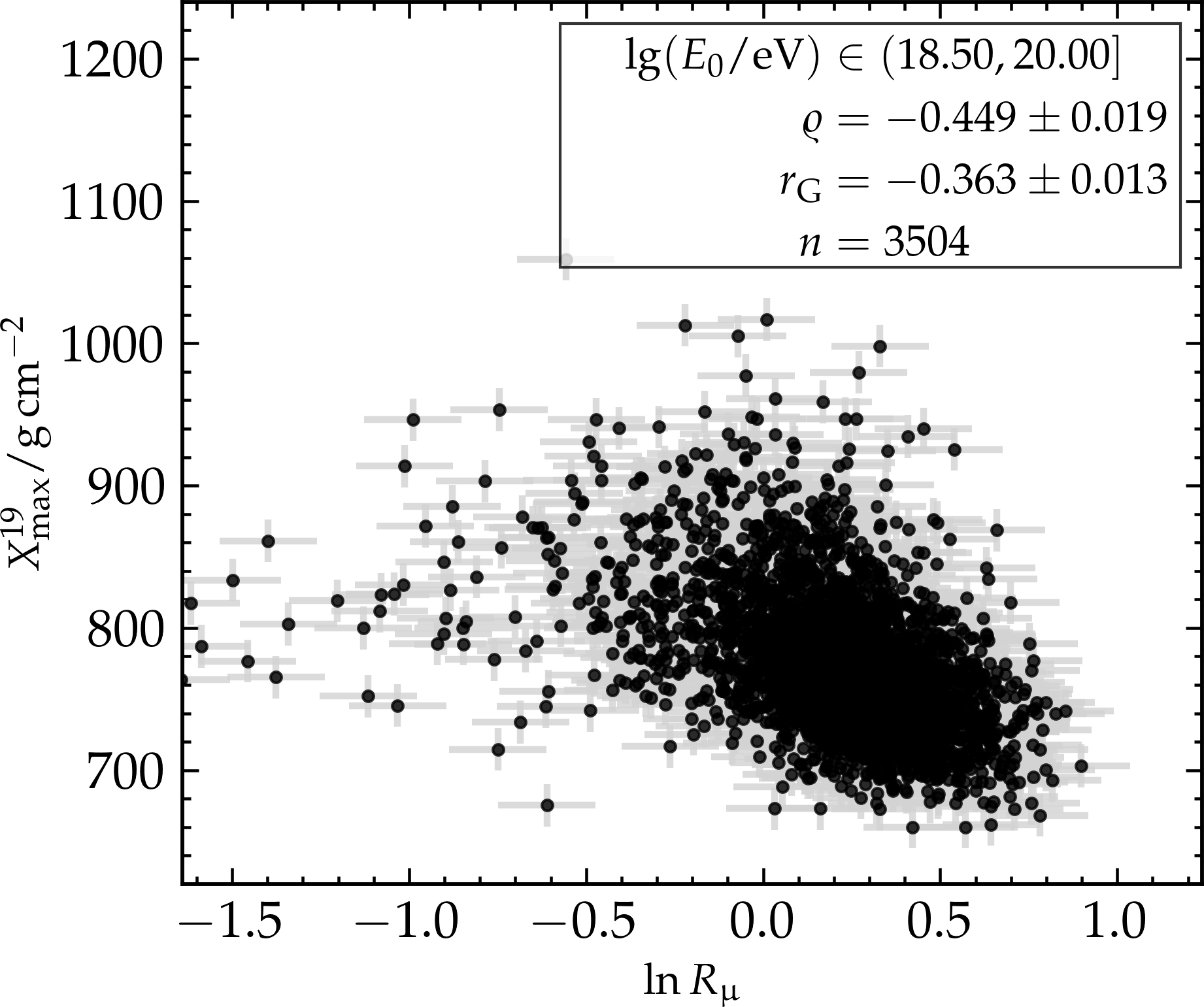}
    \put(17,15){\red{Preliminary}}
    \end{overpic}
    \caption{\emph{Left:} The relative muon number $R_\upmu$ as a function of the primary energy $E_0^\text{FD}$ measured by the fluorescence detector.
    The expectations from different hadronic interaction models and primary particles are given as blue and red reference lines.
    Expectations according to the $\Xmax$ measurements of the showers are shown in orange.
    The systematic uncertainties of the estimated number of muons are shown as black brackets around the data points.
    \emph{Right:} Event-by-event correlation of the depth of the shower maximum $X_\text{max}^{19}$ and the relative muon number $R_\upmu$, see Ref.~\cite{PierreAuger:2023vcr} for details.
    }
    \label{fig:lnrmu}
\end{figure}

The number of muons is a quantitative proxy for the hadronic particle production in the shower development and is therefore related to the nuclear mass $A$ of the initial primary cosmic ray.
For a primary particle with energy $E_0$, the number of muons produced per nucleus is not linearly proportional to the energy, but approximately follows $\propto A^{1-\beta}E_0^\beta$, with $\beta\simeq0.95$~\cite{Matthews:2005sd}.
Thus, more muons will be produced in air showers initiated by heavy nuclei,\footnote{For iron nuclei (with $A_\text{Fe}=56$) relative to proton nuclei (with $A_\text{p}=1$) we expect an increase of a factor of $N_\text{Fe}/N_\text{p}\approx56^{1-\beta}=1.2$ in muons produced, i.e.\ 20\% increase.} as the primary energy is approximately evenly distributed among the nucleons of the primary nucleus.
To accurately estimate the number of shower muons from detector data, an energy estimator that is independent of the shower particle footprint can help to disentangle the apparent dependence of the number of muons with $E_0$.
Estimating $E_0$ using the direct measurement of the shower profile by the fluorescence detector, we thus can accurately estimate the number of muons produced as long as the expected shower footprint of the electromagnetic cascade is known.
Therefore, the energy estimate $E_0^\text{FD}$ by the fluorescence detector is used as an input for the model described in \cref{sec:univ}, and the number of muons can be subsequently inferred using a fit.
This procedure and the expected performance obtained from simulations are described in detail in Ref.~\cite{PierreAuger:2023vcr}.

The relative number of muons $R_\upmu$ measured with the universality model is shown as a function of the primary energy $E_0$ in~\cref{fig:lnrmu}\,(left), alongside expectations from simulated air showers using the \textsc{Epos-LHC} and \textsc{QGSJetII-04}~\cite{Ostapchenko:2010vb} models of hadronic interaction.
Note that the data selection for the number of muons shown in~\cref{fig:lnrmu} is limited to \emph{Golden Hybrid} events where both the fluorescence and surface detector systems were detecting events at the same time; for these events, however, the correlation of the two independently-reconstructed mass-sensitive observables yields interesting insights into the composition of the cosmic-ray beam, see~\cref{fig:lnrmu}\,(right) and Ref.~\cite{PierreAuger:2023vcr}.
The relative muon number $\ln R_\upmu$ is expected to increase linearly with $\ln A$ of the primary cosmic ray.
The expectations for the number of muons from the measurements of the average $\Xmax$ are substantially lower than the data.
This tension is known as the \emph{muon deficit} or \emph{muon puzzle}, which is present in the data of the Pierre Auger Observatory and other air-shower experiments~\cite{PierreAuger:2014ucz,ArteagaVelazquez:2023fda}.

\section{Surface Detector Data: Depth of the Shower Maximum}
\label{sec:xmax}

The detector time traces (time-dependent signals from WCDs) are directly related to the shower development.
This empirically-confirmed relationship was previously already used to infer mass-composition information~\cite{PierreAuger:2017tlx} and to estimate the muon production depth~\cite{Collica:2016bck} using the SD data.
In the universality model, the \emph{time quantiles} $t_x$ of traces at $x$-fraction of total signal from the detectors are parametrized relative to the arrival time $t_\text{pf}$ of the plane shower front.
The time quantile $t_{40}$, at which a $40\%$-fraction of the (total) signal is deposited, was found to be directly related to the depth difference of the detector to the shower maximum at $X_\text{max}$~\cite{Stadelmaier:2024pae}.
Using a quasi-spherical shower model, $t_{40}$ is expressed as a function of the shower geometry for each detector station; for details, again see Ref.~\cite{Stadelmaier:2024pae}.

\begin{figure}
    \centering
    \def\h{0.37}
    \includegraphics[height=\h\textwidth]{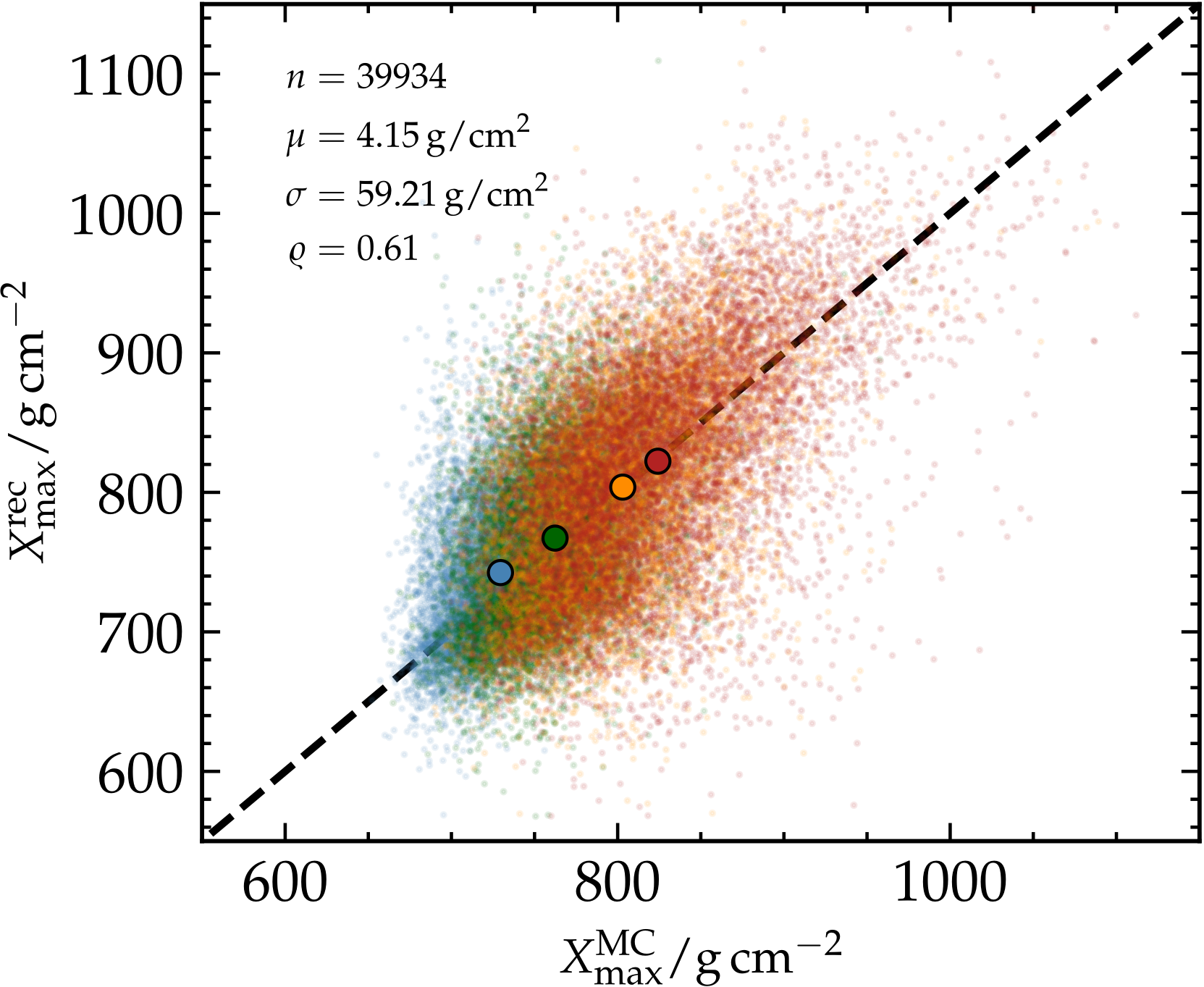}
    \hfill
    \includegraphics[height=\h\textwidth]{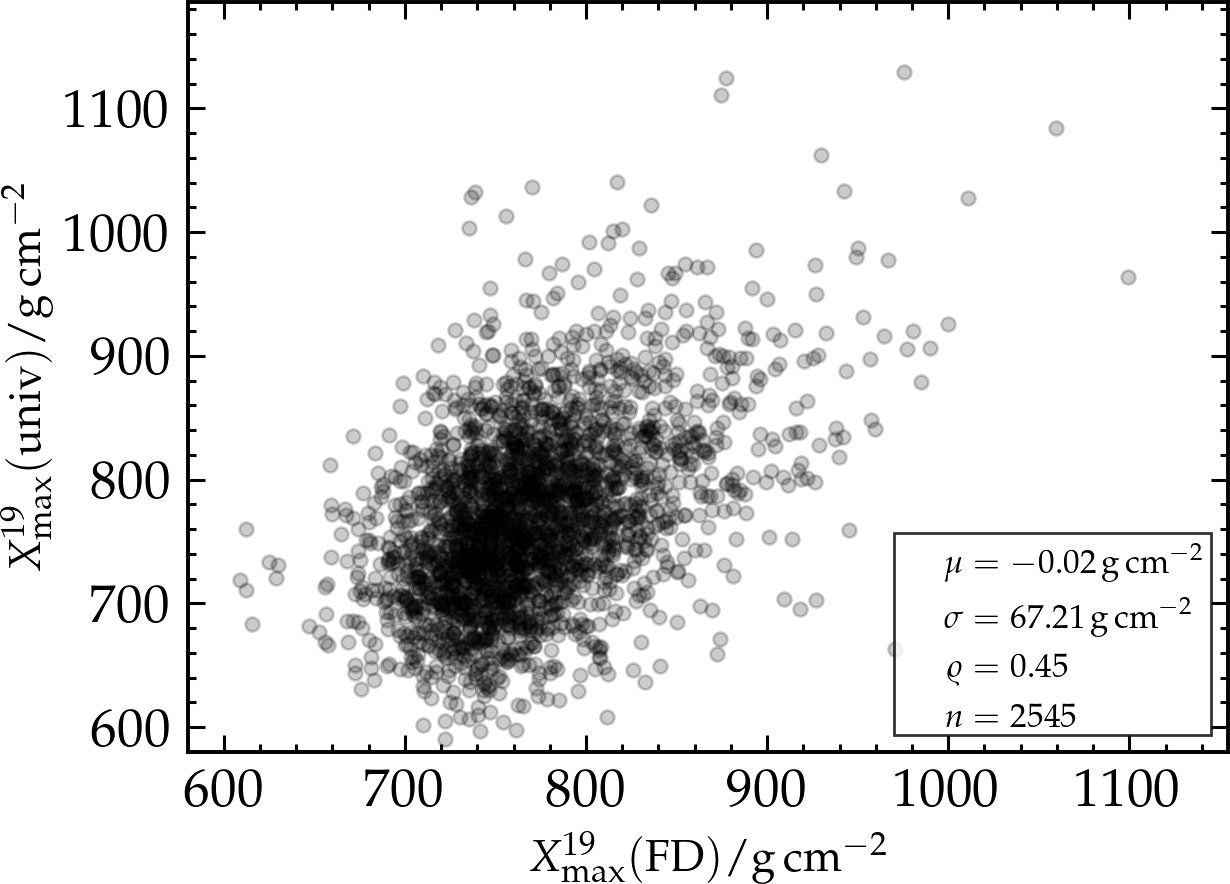}
    \caption{\emph{Left:} True and reconstructed values of $\Xmax$ for simulated showers above a primary energy of $3\,\EeV$.
    The number of data points as well as overall mean and width of the residuals $\Delta\Xmax^\text{rec}$ are given alongside the Pearson correlation coefficient in the upper left corner.
    Dot colors correspond to {\color{red}proton}, {\color{orange}helium}, {\color{ForestGreen}oxygen}, and {\color{Cerulean}iron}.
    \emph{Right:} Correlation of the estimated (univ) and directly measured (FD) values of the depth of the shower maxima in the Golden Hybrid data set.
    The moments of the residual distribution as well as the Pearson correlation and the number of events is given in the legend.
    See the text for details.}
    \label{fig:xmaxxmax}
\end{figure}

Using this model, we fit the time-dependent SD data to estimate $\Xmax$.
For this purpose the traces are first normalized (i.e.\ treated as a PDF and then using a corresponding CDF for its time quantiles) to reduce the effect of both the number of muons and the primary energy on signal size to influence the $\Xmax$ fit.
We thus do not expect the results to be artificially correlated with any other observables.
Stations that are far away from the shower axis (${\geq}1800$\,m) are removed from the fit, since their traces carry little to no information about $X_\text{max}$, and the model fails to accurately describe the shower data at distances beyond ${\simeq}2000$\,m.
The performance of this indirect reconstruction method to estimate $X_\text{max}$ was tested both using simulations and Golden Hybrid data.
For the latter, the SD data was analyzed independently of the FD information and the estimated values of $X_\text{max}$ were compared for each event.
\cref{fig:xmaxxmax} shows the event-level validation of the method for both simulations and data using directly and indirectly obtained values of $X_\text{max}$.
Since the energy range in \cref{fig:xmaxxmax} spans more than one order of magnitude, a constant elongation rate was removed from the $X_\text{max}$ values\footnote{where $X_\text{max}^{19}=X_\text{max}-D\,\lg(E_0/10^{19}\,\eV)$ using a constant decadal elongation rate, $D\simeq 56\,\gcm$.} to obtain the $X_\text{max}^{19}$ references.
Although the performance of the universality reconstruction is weaker when compared to the novel machine-learning methods~\cite{Hahn:2025ybh,PierreAuger:2023tyq,PierreAuger:2024flk,PierreAuger:2024nzw}, we still observe a significant correlation of the universality and FD data sets.
The reconstruction using the universality model is therefore confirmed to be able to estimate $X_\text{max}$ from the detector time traces.

\begin{figure}
    \centering
    \def\h{0.35}
    \begin{overpic}[height=\h\textwidth]{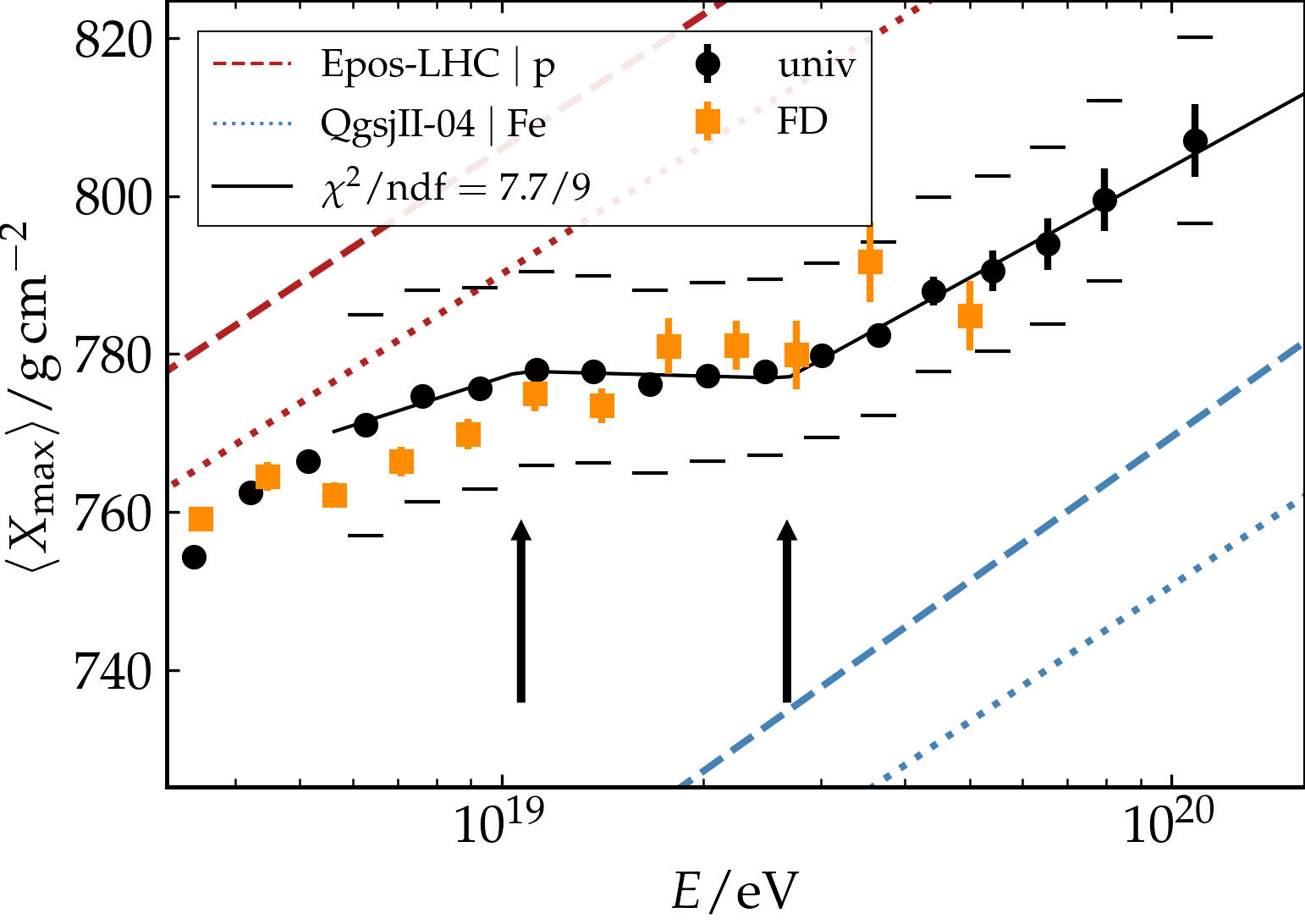}
    \put(72,14){\red{Preliminary}}
    \end{overpic}\hfill
    \begin{overpic}[height=\h\textwidth]{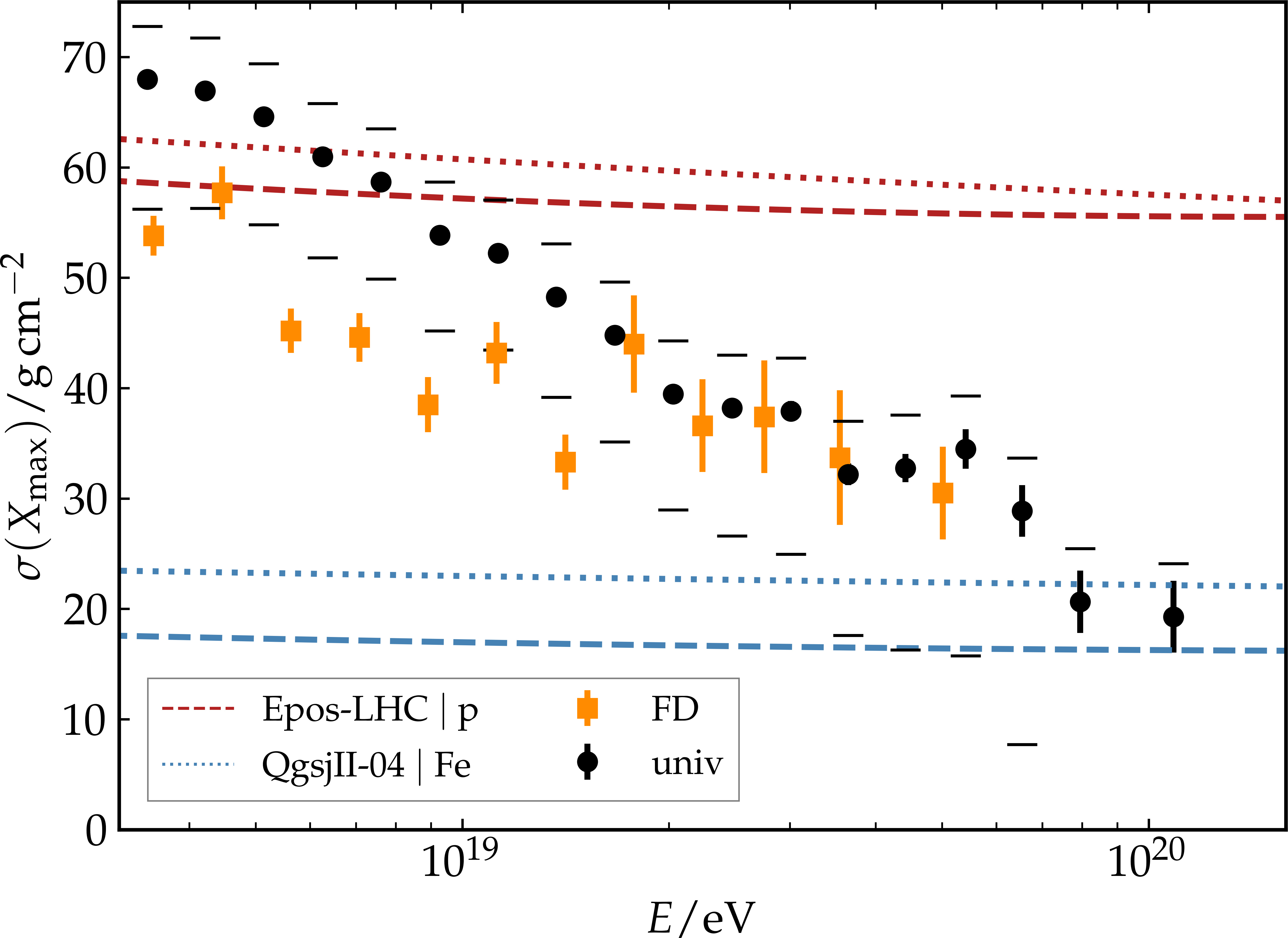}
    \put(72,12){\red{Preliminary}}
    \end{overpic}
    \caption{\emph{Left:} $\avg{X_\text{max}}$ as a function of the (SD reconstructed) energy with a broken-line fit.
    \emph{Right:} Fluctuations of $\Xmax$ given as the estimated standard deviation of the $\Xmax$-distribution, $\sigma(\Xmax)$, as a function of the primary energy.
    The black markers denote the mean (or standard-deviation) values from the universality fit, orange markers show the results from direct FD measurements.
    Black caps stand for the systematic uncertainties.
    Red and blue dashed and dotted lines indicate the expectation values from simulations using the \textsc{Epos-LHC} and \textsc{QGSJetII-04} models of hadronic interactions for showers from proton and iron primary particles, respectively.}
    \label{fig:results}
\end{figure}

The $\Xmax$ as estimated by the universality-model fit applied to the Phase-I SD data of the Pierre Auger Observatory are shown in \cref{fig:results}.
The average $X_\text{max}$ as a function of the primary energy $E$ is monotonically increasing.
However, compared to expectations from simulations, the corresponding mean mass $\langle\ln A\rangle$ of UHECRs is increasing with $E$.
Furthermore, $\avg{X_\text{max}}$ at high energies evolves with the same elongation rate as expected from a pure-primary beam, which implies a somewhat constant mass composition above ${\simeq}30$\,EeV.
Where they are available, the results of the universality-model fit (after calibration of the mean) are in reasonable agreement with the data from direct FD measurements in terms of the evolution with primary energy.
For the first three data points in \cref{fig:results}\,(left) we did not assign systematic uncertainties, because these energies were omitted from the calibration fit, see~\cref{sec:sys}.

The fluctuations of $X_\text{max}$ as a function of the primary energy $E$ were obtained by removing the expected intrinsic precision, as estimated using Monte-Carlo simulations.
Above 10\,EeV, the qualitative behavior of $\sigma(X_\text{max})$ as a function of $E$ agrees well with the FD data and with expectations from other methods~\cite{PierreAuger:2024nzw,PierreAuger:2024flk,PierreAuger:2017tlx}.
At lower energies ($E_0\lesssim10$\,EeV), the universality results consistently show larger fluctuations than expected.
This might be due to discrepancies of the performance of the method on measured data with respect to simulations, or possibly due to individual outliers being over-represented.
Nevertheless, the behavior of $\sigma(X_
\text{max})$ as a function of $E$ implies a heavier and more pure mass composition at the highest energies, and is compatible with a proton-dominated (or helium-dominated) composition close to the ankle region around ${\simeq}3\,\EeV$ for the \textsc{QGSJetII-04} (or \textsc{Epos-LHC}) model.

Taking a closer look at the evolution of $\Xmax$ as a function of primary energy above $E_0 = 10^{18.8}\,\eV \simeq 6\,\EeV$, it is immediately clear that the elongation rate appears not to be constant.
This has already been studied in depth in Refs.~\cite{PierreAuger:2024nzw,PierreAuger:2024flk}, where the elongation rate is best fit with a broken line using two breaks.
For comparison, a fit to the data shown in \cref{fig:results} using only one break yields $\chi^2/\text{ndf} \simeq 45.2 / 11$ corresponding to a very low probability $p(\chi^2(\text{ndf})) \simeq 3{\times}10^{-6}$.

\section{Systematic Uncertainties and Calibration}
\label{sec:sys}

Since the results obtained when applying the universality model to data rely heavily on the FD either as a source of calibration or direct input, systematic uncertainties are inherited as well.
The systematic uncertainties displayed in \cref{fig:lnrmu,fig:results} are mostly a direct result of the ${\simeq}14\%$ systematic uncertainty of the energy scale of FD and the uncertainty of the $X_\text{max}$ calibration.
The latter is necessary to correct for an overall difference of the mean $X_\text{max}$ provided by raw results of the universality fit and direct measurements; a similar calibration was performed in Refs.~\cite{PierreAuger:2017tlx,PierreAuger:2021fkf,PierreAuger:2024nzw}.

\section{Discussion and Summary}

The universality-based shower model, described in this work, is an attempt to describe the generalized shower development and use it to reconstruct shower observables.
Depending on the input parameters, the model can be used to estimate the number of muons produced in the shower, and/or the depth of the shower maximum on an event level.
In general, the reconstructed number of muons is more accurate if an independent energy estimator is used; when trying to reconstruct $\Xmax$ from the time-dependent signal traces, no event-level fluorescence detector data is required.

The results of the reconstructed shower observables can be used either to infer the details of hadronic interactions in the shower development and maybe introduce new perspectives to mass-composition analyses (see e.g.\ Ref.~\cite{Vicha:2025fxx}), or to qualitatively identify and separate \emph{lighter} or \emph{heavier} events in the data of the Pierre Auger Observatory.
%\cref{fig:spec} shows the number of reconstructed events as a function of energy above 8\,EeV where an example 30\% quantile cut was applied to the overall $\Xmax$ data to separate \emph{light} and \emph{heavy} cosmic rays.
Such separated data sets can be used, for example, to reduce background when conducting arrival-direction analyses (see e.g.\ Ref.~\cite{PierreAuger:2025apollonio}).

%\begin{figure}
%    \centering
%    \includegraphics[height=\gHeight\textwidth]{lollo2-pseudo_spectrum}
%    \caption{\emph{``Pseudo'' spectrum}, number of events in equal bins of logarithmic reconstructed primary energy for different mass-composition species of the data of the Pierre Auger Observatory.
%    Error bands indicate the Poissonian uncertainty of number of events per bin, diagonal lines indicate the number of events per bin.}
%    \label{fig:spec}
%\end{figure}

\let\oldbibliography\thebibliography
\renewcommand{\thebibliography}[1]{%
  \oldbibliography{#1}%
  \setlength{\itemsep}{0pt}%
}

{\small
\bibliography{bibfile}{}
\bibliographystyle{JHEP_mod}
}

\clearpage

\section*{The Pierre Auger Collaboration}

{\footnotesize\setlength{\baselineskip}{10pt}
\noindent
\begin{wrapfigure}[11]{l}{0.12\linewidth}
\vspace{-4pt}
\href{https://www.auger.org}{\includegraphics[width=0.98\linewidth]{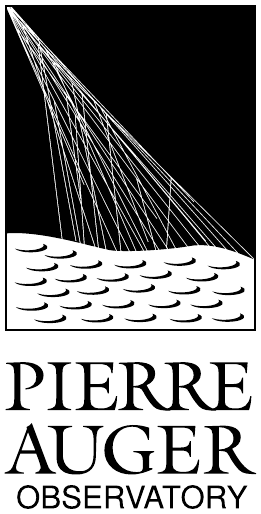}}
\end{wrapfigure}
\begin{sloppypar}\noindent
% created on 2025-06-06
A.~Abdul Halim$^{13}$,
P.~Abreu$^{70}$,
M.~Aglietta$^{53,51}$,
I.~Allekotte$^{1}$,
K.~Almeida Cheminant$^{78,77}$,
A.~Almela$^{7,12}$,
R.~Aloisio$^{44,45}$,
J.~Alvarez-Mu\~niz$^{76}$,
A.~Ambrosone$^{44}$,
J.~Ammerman Yebra$^{76}$
% ...

\end{sloppypar}
\begin{center}
\end{center}

\vspace{1ex}
% created on 2025-06-06
% needs \usepackage{enumitem}
\begin{description}[labelsep=0.2em,align=right,labelwidth=0.7em,labelindent=0em,leftmargin=2em,noitemsep,before={\renewcommand\makelabel[1]{##1 }}]
\item[$^{1}$] Centro At\'omico Bariloche and Instituto Balseiro (CNEA-UNCuyo-CONICET), San Carlos de Bariloche, Argentina
\item[$^{2}$] Departamento de F\'\i{}sica and Departamento de Ciencias de la Atm\'osfera y los Oc\'eanos, FCEyN, Universidad de Buenos Aires and CONICET, Buenos Aires, Argentina
\item[$^{3}$] IFLP, Universidad Nacional de La Plata and CONICET, La Plata, Argentina
\item[$^{4}$] Instituto de Astronom\'\i{}a y F\'\i{}sica del Espacio (IAFE, CONICET-UBA), Buenos Aires, Argentina
\item[$^{5}$] Instituto de F\'\i{}sica de Rosario (IFIR) -- CONICET/U.N.R.\ and Facultad de Ciencias Bioqu\'\i{}micas y Farmac\'euticas U.N.R., Rosario, Argentina
\item[$^{6}$] Instituto de Tecnolog\'\i{}as en Detecci\'on y Astropart\'\i{}culas (CNEA, CONICET, UNSAM), and Universidad Tecnol\'ogica Nacional -- Facultad Regional Mendoza (CONICET/CNEA), Mendoza, Argentina
\item[$^{7}$] Instituto de Tecnolog\'\i{}as en Detecci\'on y Astropart\'\i{}culas (CNEA, CONICET, UNSAM), Buenos Aires, Argentina
\item[$^{8}$] International Center of Advanced Studies and Instituto de Ciencias F\'\i{}sicas, ECyT-UNSAM and CONICET, Campus Miguelete -- San Mart\'\i{}n, Buenos Aires, Argentina
\item[$^{9}$] Laboratorio Atm\'osfera -- Departamento de Investigaciones en L\'aseres y sus Aplicaciones -- UNIDEF (CITEDEF-CONICET), Argentina
\item[$^{10}$] Observatorio Pierre Auger, Malarg\"ue, Argentina
\item[$^{11}$] Observatorio Pierre Auger and Comisi\'on Nacional de Energ\'\i{}a At\'omica, Malarg\"ue, Argentina
\item[$^{12}$] Universidad Tecnol\'ogica Nacional -- Facultad Regional Buenos Aires, Buenos Aires, Argentina
\item[$^{13}$] University of Adelaide, Adelaide, S.A., Australia
\item[$^{14}$] Universit\'e Libre de Bruxelles (ULB), Brussels, Belgium
\item[$^{15}$] Vrije Universiteit Brussels, Brussels, Belgium
\item[$^{16}$] Centro Brasileiro de Pesquisas Fisicas, Rio de Janeiro, RJ, Brazil
\item[$^{17}$] Centro Federal de Educa\c{c}\~ao Tecnol\'ogica Celso Suckow da Fonseca, Petropolis, Brazil
\item[$^{18}$] Instituto Federal de Educa\c{c}\~ao, Ci\^encia e Tecnologia do Rio de Janeiro (IFRJ), Brazil
\item[$^{19}$] Universidade de S\~ao Paulo, Escola de Engenharia de Lorena, Lorena, SP, Brazil
\item[$^{20}$] Universidade de S\~ao Paulo, Instituto de F\'\i{}sica de S\~ao Carlos, S\~ao Carlos, SP, Brazil
\item[$^{21}$] Universidade de S\~ao Paulo, Instituto de F\'\i{}sica, S\~ao Paulo, SP, Brazil
\item[$^{22}$] Universidade Estadual de Campinas (UNICAMP), IFGW, Campinas, SP, Brazil
\item[$^{23}$] Universidade Estadual de Feira de Santana, Feira de Santana, Brazil
\item[$^{24}$] Universidade Federal de Campina Grande, Centro de Ciencias e Tecnologia, Campina Grande, Brazil
\item[$^{25}$] Universidade Federal do ABC, Santo Andr\'e, SP, Brazil
\item[$^{26}$] Universidade Federal do Paran\'a, Setor Palotina, Palotina, Brazil
\item[$^{27}$] Universidade Federal do Rio de Janeiro, Instituto de F\'\i{}sica, Rio de Janeiro, RJ, Brazil
\item[$^{28}$] Universidad de Medell\'\i{}n, Medell\'\i{}n, Colombia
\item[$^{29}$] Universidad Industrial de Santander, Bucaramanga, Colombia
\item[$^{30}$] Charles University, Faculty of Mathematics and Physics, Institute of Particle and Nuclear Physics, Prague, Czech Republic
\item[$^{31}$] Institute of Physics of the Czech Academy of Sciences, Prague, Czech Republic
\item[$^{32}$] Palacky University, Olomouc, Czech Republic
\item[$^{33}$] CNRS/IN2P3, IJCLab, Universit\'e Paris-Saclay, Orsay, France
\item[$^{34}$] Laboratoire de Physique Nucl\'eaire et de Hautes Energies (LPNHE), Sorbonne Universit\'e, Universit\'e de Paris, CNRS-IN2P3, Paris, France
\item[$^{35}$] Univ.\ Grenoble Alpes, CNRS, Grenoble Institute of Engineering Univ.\ Grenoble Alpes, LPSC-IN2P3, 38000 Grenoble, France
\item[$^{36}$] Universit\'e Paris-Saclay, CNRS/IN2P3, IJCLab, Orsay, France
\item[$^{37}$] Bergische Universit\"at Wuppertal, Department of Physics, Wuppertal, Germany
\item[$^{38}$] Karlsruhe Institute of Technology (KIT), Institute for Experimental Particle Physics, Karlsruhe, Germany
\item[$^{39}$] Karlsruhe Institute of Technology (KIT), Institut f\"ur Prozessdatenverarbeitung und Elektronik, Karlsruhe, Germany
\item[$^{40}$] Karlsruhe Institute of Technology (KIT), Institute for Astroparticle Physics, Karlsruhe, Germany
\item[$^{41}$] RWTH Aachen University, III.\ Physikalisches Institut A, Aachen, Germany
\item[$^{42}$] Universit\"at Hamburg, II.\ Institut f\"ur Theoretische Physik, Hamburg, Germany
\item[$^{43}$] Universit\"at Siegen, Department Physik -- Experimentelle Teilchenphysik, Siegen, Germany
\item[$^{44}$] Gran Sasso Science Institute, L'Aquila, Italy
\item[$^{45}$] INFN Laboratori Nazionali del Gran Sasso, Assergi (L'Aquila), Italy
\item[$^{46}$] INFN, Sezione di Catania, Catania, Italy
\item[$^{47}$] INFN, Sezione di Lecce, Lecce, Italy
\item[$^{48}$] INFN, Sezione di Milano, Milano, Italy
\item[$^{49}$] INFN, Sezione di Napoli, Napoli, Italy
\item[$^{50}$] INFN, Sezione di Roma ``Tor Vergata'', Roma, Italy
\item[$^{51}$] INFN, Sezione di Torino, Torino, Italy
\item[$^{52}$] Istituto di Astrofisica Spaziale e Fisica Cosmica di Palermo (INAF), Palermo, Italy
\item[$^{53}$] Osservatorio Astrofisico di Torino (INAF), Torino, Italy
\item[$^{54}$] Politecnico di Milano, Dipartimento di Scienze e Tecnologie Aerospaziali , Milano, Italy
\item[$^{55}$] Universit\`a del Salento, Dipartimento di Matematica e Fisica ``E.\ De Giorgi'', Lecce, Italy
\item[$^{56}$] Universit\`a dell'Aquila, Dipartimento di Scienze Fisiche e Chimiche, L'Aquila, Italy
\item[$^{57}$] Universit\`a di Catania, Dipartimento di Fisica e Astronomia ``Ettore Majorana``, Catania, Italy
\item[$^{58}$] Universit\`a di Milano, Dipartimento di Fisica, Milano, Italy
\item[$^{59}$] Universit\`a di Napoli ``Federico II'', Dipartimento di Fisica ``Ettore Pancini'', Napoli, Italy
\item[$^{60}$] Universit\`a di Palermo, Dipartimento di Fisica e Chimica ''E.\ Segr\`e'', Palermo, Italy
\item[$^{61}$] Universit\`a di Roma ``Tor Vergata'', Dipartimento di Fisica, Roma, Italy
\item[$^{62}$] Universit\`a Torino, Dipartimento di Fisica, Torino, Italy
\item[$^{63}$] Benem\'erita Universidad Aut\'onoma de Puebla, Puebla, M\'exico
\item[$^{64}$] Unidad Profesional Interdisciplinaria en Ingenier\'\i{}a y Tecnolog\'\i{}as Avanzadas del Instituto Polit\'ecnico Nacional (UPIITA-IPN), M\'exico, D.F., M\'exico
\item[$^{65}$] Universidad Aut\'onoma de Chiapas, Tuxtla Guti\'errez, Chiapas, M\'exico
\item[$^{66}$] Universidad Michoacana de San Nicol\'as de Hidalgo, Morelia, Michoac\'an, M\'exico
\item[$^{67}$] Universidad Nacional Aut\'onoma de M\'exico, M\'exico, D.F., M\'exico
\item[$^{68}$] Institute of Nuclear Physics PAN, Krakow, Poland
\item[$^{69}$] University of \L{}\'od\'z, Faculty of High-Energy Astrophysics,\L{}\'od\'z, Poland
\item[$^{70}$] Laborat\'orio de Instrumenta\c{c}\~ao e F\'\i{}sica Experimental de Part\'\i{}culas -- LIP and Instituto Superior T\'ecnico -- IST, Universidade de Lisboa -- UL, Lisboa, Portugal
\item[$^{71}$] ``Horia Hulubei'' National Institute for Physics and Nuclear Engineering, Bucharest-Magurele, Romania
\item[$^{72}$] Institute of Space Science, Bucharest-Magurele, Romania
\item[$^{73}$] Center for Astrophysics and Cosmology (CAC), University of Nova Gorica, Nova Gorica, Slovenia
\item[$^{74}$] Experimental Particle Physics Department, J.\ Stefan Institute, Ljubljana, Slovenia
\item[$^{75}$] Universidad de Granada and C.A.F.P.E., Granada, Spain
\item[$^{76}$] Instituto Galego de F\'\i{}sica de Altas Enerx\'\i{}as (IGFAE), Universidade de Santiago de Compostela, Santiago de Compostela, Spain
\item[$^{77}$] IMAPP, Radboud University Nijmegen, Nijmegen, The Netherlands
\item[$^{78}$] Nationaal Instituut voor Kernfysica en Hoge Energie Fysica (NIKHEF), Science Park, Amsterdam, The Netherlands
\item[$^{79}$] Stichting Astronomisch Onderzoek in Nederland (ASTRON), Dwingeloo, The Netherlands
\item[$^{80}$] Universiteit van Amsterdam, Faculty of Science, Amsterdam, The Netherlands
\item[$^{81}$] Case Western Reserve University, Cleveland, OH, USA
\item[$^{82}$] Colorado School of Mines, Golden, CO, USA
\item[$^{83}$] Department of Physics and Astronomy, Lehman College, City University of New York, Bronx, NY, USA
\item[$^{84}$] Michigan Technological University, Houghton, MI, USA
\item[$^{85}$] New York University, New York, NY, USA
\item[$^{86}$] University of Chicago, Enrico Fermi Institute, Chicago, IL, USA
\item[$^{87}$] University of Delaware, Department of Physics and Astronomy, Bartol Research Institute, Newark, DE, USA
\item[] -----
\item[$^{a}$] Max-Planck-Institut f\"ur Radioastronomie, Bonn, Germany
\item[$^{b}$] also at Kapteyn Institute, University of Groningen, Groningen, The Netherlands
\item[$^{c}$] School of Physics and Astronomy, University of Leeds, Leeds, United Kingdom
\item[$^{d}$] Fermi National Accelerator Laboratory, Fermilab, Batavia, IL, USA
\item[$^{e}$] Pennsylvania State University, University Park, PA, USA
\item[$^{f}$] Colorado State University, Fort Collins, CO, USA
\item[$^{g}$] Louisiana State University, Baton Rouge, LA, USA
\item[$^{h}$] now at Graduate School of Science, Osaka Metropolitan University, Osaka, Japan
\item[$^{i}$] Institut universitaire de France (IUF), France
\item[$^{j}$] now at Technische Universit\"at Dortmund and Ruhr-Universit\"at Bochum, Dortmund and Bochum, Germany
\end{description}

\clearpage

% created on 2025-06-06
\section*{Acknowledgments}

\begin{sloppypar}
The successful installation, commissioning, and operation of the Pierre
Auger Observatory would not have been possible without the strong
commitment and effort from the technical and administrative staff in
Malarg\"ue. We are very grateful to the following agencies and
organizations for financial support:
\end{sloppypar}

\begin{sloppypar}
Argentina -- Comisi\'on Nacional de Energ\'\i{}a At\'omica; Agencia Nacional de
Promoci\'on Cient\'\i{}fica y Tecnol\'ogica (ANPCyT); Consejo Nacional de
Investigaciones Cient\'\i{}ficas y T\'ecnicas (CONICET); Gobierno de la
Provincia de Mendoza; Municipalidad de Malarg\"ue; NDM Holdings and Valle
Las Le\~nas; in gratitude for their continuing cooperation over land
access; Australia -- the Australian Research Council; Belgium -- Fonds
de la Recherche Scientifique (FNRS); Research Foundation Flanders (FWO),
Marie Curie Action of the European Union Grant No.~101107047; Brazil --
Conselho Nacional de Desenvolvimento Cient\'\i{}fico e Tecnol\'ogico (CNPq);
Financiadora de Estudos e Projetos (FINEP); Funda\c{c}\~ao de Amparo \`a
Pesquisa do Estado de Rio de Janeiro (FAPERJ); S\~ao Paulo Research
Foundation (FAPESP) Grants No.~2019/10151-2, No.~2010/07359-6 and
No.~1999/05404-3; Minist\'erio da Ci\^encia, Tecnologia, Inova\c{c}\~oes e
Comunica\c{c}\~oes (MCTIC); Czech Republic -- GACR 24-13049S, CAS LQ100102401,
MEYS LM2023032, CZ.02.1.01/0.0/0.0/16{\textunderscore}013/0001402,
CZ.02.1.01/0.0/0.0/18{\textunderscore}046/0016010 and
CZ.02.1.01/0.0/0.0/17{\textunderscore}049/0008422 and CZ.02.01.01/00/22{\textunderscore}008/0004632;
France -- Centre de Calcul IN2P3/CNRS; Centre National de la Recherche
Scientifique (CNRS); Conseil R\'egional Ile-de-France; D\'epartement
Physique Nucl\'eaire et Corpusculaire (PNC-IN2P3/CNRS); D\'epartement
Sciences de l'Univers (SDU-INSU/CNRS); Institut Lagrange de Paris (ILP)
Grant No.~LABEX ANR-10-LABX-63 within the Investissements d'Avenir
Programme Grant No.~ANR-11-IDEX-0004-02; Germany -- Bundesministerium
f\"ur Bildung und Forschung (BMBF); Deutsche Forschungsgemeinschaft (DFG);
Finanzministerium Baden-W\"urttemberg; Helmholtz Alliance for
Astroparticle Physics (HAP); Helmholtz-Gemeinschaft Deutscher
Forschungszentren (HGF); Ministerium f\"ur Kultur und Wissenschaft des
Landes Nordrhein-Westfalen; Ministerium f\"ur Wissenschaft, Forschung und
Kunst des Landes Baden-W\"urttemberg; Italy -- Istituto Nazionale di
Fisica Nucleare (INFN); Istituto Nazionale di Astrofisica (INAF);
Ministero dell'Universit\`a e della Ricerca (MUR); CETEMPS Center of
Excellence; Ministero degli Affari Esteri (MAE), ICSC Centro Nazionale
di Ricerca in High Performance Computing, Big Data and Quantum
Computing, funded by European Union NextGenerationEU, reference code
CN{\textunderscore}00000013; M\'exico -- Consejo Nacional de Ciencia y Tecnolog\'\i{}a
(CONACYT) No.~167733; Universidad Nacional Aut\'onoma de M\'exico (UNAM);
PAPIIT DGAPA-UNAM; The Netherlands -- Ministry of Education, Culture and
Science; Netherlands Organisation for Scientific Research (NWO); Dutch
national e-infrastructure with the support of SURF Cooperative; Poland
-- Ministry of Education and Science, grants No.~DIR/WK/2018/11 and
2022/WK/12; National Science Centre, grants No.~2016/22/M/ST9/00198,
2016/23/B/ST9/01635, 2020/39/B/ST9/01398, and 2022/45/B/ST9/02163;
Portugal -- Portuguese national funds and FEDER funds within Programa
Operacional Factores de Competitividade through Funda\c{c}\~ao para a Ci\^encia
e a Tecnologia (COMPETE); Romania -- Ministry of Research, Innovation
and Digitization, CNCS-UEFISCDI, contract no.~30N/2023 under Romanian
National Core Program LAPLAS VII, grant no.~PN 23 21 01 02 and project
number PN-III-P1-1.1-TE-2021-0924/TE57/2022, within PNCDI III; Slovenia
-- Slovenian Research Agency, grants P1-0031, P1-0385, I0-0033, N1-0111;
Spain -- Ministerio de Ciencia e Innovaci\'on/Agencia Estatal de
Investigaci\'on (PID2019-105544GB-I00, PID2022-140510NB-I00 and
RYC2019-027017-I), Xunta de Galicia (CIGUS Network of Research Centers,
Consolidaci\'on 2021 GRC GI-2033, ED431C-2021/22 and ED431F-2022/15),
Junta de Andaluc\'\i{}a (SOMM17/6104/UGR and P18-FR-4314), and the European
Union (Marie Sklodowska-Curie 101065027 and ERDF); USA -- Department of
Energy, Contracts No.~DE-AC02-07CH11359, No.~DE-FR02-04ER41300,
No.~DE-FG02-99ER41107 and No.~DE-SC0011689; National Science Foundation,
Grant No.~0450696, and NSF-2013199; The Grainger Foundation; Marie
Curie-IRSES/EPLANET; European Particle Physics Latin American Network;
and UNESCO.
\end{sloppypar}

}

\end{document}